\documentclass[twocolumn,aps,floats,showpacs,amsmath]{revtex4}
\usepackage{epsfig}
\usepackage{amssymb}
\begin{document} 

\title{Hybrid soft-mode and off-center Ti model of barium titanate}

\author{R. Pirc and R. Blinc}

\affiliation{Jo\v zef Stefan Institute,  P.O. Box 3000, 1001 Ljubljana, 
Slovenia}

\date{\today}

\begin{abstract}

It has been recently established by NMR techniques that in the high 
temperature cubic phase of BaTiO$_3$ the Ti ions are not confined to 
the high symmetry cubic sites, but rather occupy one of the eight 
off-center positions along the $[111]$ directions. The off-center Ti 
picture is in apparent contrast with most soft-mode type theoretical 
descriptions of this classical perovskite ferroelectric. Here we apply 
a mesoscopic model of BaTiO$_3$, assuming that the symmetrized occupation 
operators for the Ti off-center sites are linearly coupled to the normal 
coordinates for lattice vibrations. On the time scale of Ti intersite jumps,
most phonon modes are fast and thus merely contribute 
to an effective static Ti-Ti interaction. Close to the stability limit 
for the soft TO optic modes, however, the phonon time scale becomes 
comparable to the relaxation time for the Ti occupational states of 
$T_{1u}$ symmetry, and a hybrid vibrational-orientational soft mode appears. 
The frequency of the hybrid soft mode is calculated as a function of 
temperature and coupling strength, and its its role in the ferroelectric 
phase transition is discussed. 

\end{abstract}
\maketitle
\section{Introduction}

Barium titanate is a typical representative of perovskite ferroelectrics,  
which undergoes a cubic-to-tetragonal structural phase 
transition at the Curie temperature $T_C \simeq 403$ K. The soft-mode 
nature of this transition has been determined by neutron scattering 
\cite{S1} as well as by hyper-Raman scattering \cite{V1}, however, 
some open questions concerning the role played by the Ti ions have so far 
remained unanswered. Chaves et al. \cite{C1} proposed a thermodynamic 
model following the assumption of Comes et al. \cite{C2} that the Ti ion 
occupies one of the eight equivalent off-center sites along the 
$[111]$ directions, which predicts a nonzero microscopic dipole 
moment of each unit cell and hence a transition of an order-disorder type. 

The off-center displacements of Ti ions in the high-temperature cubic
phase of BaTiO$_3$ have recently been confirmed by NMR experiments \cite{Z1}, 
which revealed the order-disorder dynamics of Ti ions to be coexisting 
with the observable displacive features of the TO soft mode. This 
immediately raises the question about the appropriate theoretical model 
for BaTiO$_3$ and related systems, as the widely accepted soft-mode 
description implies a central position of the Ti ion in the high temperature 
cubic phase, leading to a phase transition of a purely displacive type. 
Also, first-principles calculations of the electronic structure \cite{C3} 
do not support the idea that the ferroelectric distortion is due to the Ti 
ion `rattling' in the oxygen cage. Some general features of the two types 
of phase transition have been discussed by Aubry \cite{A1} on the basis of 
a linear coupled double-well model. The crossover between the order-disorder 
and displacive transition was investigated by means of molecular dynamics 
calculations by Stachiotti et al. \cite{S2} who used a two-dimensional 
shell model of oxide perovskites. 
A microscopic tretament of perovskite ferroelectrics has been 
carried out by Girshberg and Yacoby (GY) \cite{G1,G2}, who describe 
the degrees of freedom associated with the off-center displacements 
in terms of Ising pseudospin operators and introduce a linear 
pseudospin-phonon coupling. They derive an effective coupling between 
the off-center ions, which leads to a pseudospin ordering transition.
The corresponding transition temperature is shifted from the 
instability temperature of the TO soft mode.

In this paper, we make an attempt to develop a mesoscopic treatment which 
combines the order-disorder and displacive features of the phase transition. 
We introduce a simple mechanism at the mesoscopic level, which is 
closely related to the above GY model, i.e., we adopt the linear
coupling between the symmetrized occupation operators for the off-center 
Ti sites and the normal coordinates for lattice displacements.
Assuming that on the time scale of Ti intersite jumps most phonon modes 
are fast, we show that these modes essentially contribute to a
renormalization of the static Ti-Ti interactions, which can be either 
ferro- or antiferro-distortive. Meanwhile, close to the Ti ordering transition 
the time scale of the soft TO mode becomes asymptotically comparable to the 
characteristic time for the relaxation of the Ti dipole moment. Thus
a new vibrational-orientational hybrid mode appears, which retains some 
of its original soft-mode character, however, its frequency and width is 
determined by the the relaxational dynamics of the Ti subsystem. In addition 
to the hybrid mode, a purely relaxational mode describing the intersite jumps 
of the Ti ions exists. It should be noted that the unperturbed TO soft mode
describes the anharmonic vibrations of all the ions, and thus also include 
the oscillations of Ti ions around their average equilibrium positions. 
Apart from the time scales involved, the conditions for the formation of 
a hybrid soft mode may also depend on the strength of the Ti-phonon 
coupling, which thus plays the role of a control parameter in the present
model.

\section{Coupled titanium-phonon system: Statics}

Following Chaves et al. \cite{C1} we define the occupation probabilities 
$n_{il}= \{1,0\}$ for the off-center Ti sites in the $i$-th unit cell,
where $l=1,2,\cdots ,8$ according to the convention specified in Ref.~\cite{C1}. 
Obviously, $\sum_l n_{il} = 1$. Next, we introduce a set of symmetry adapted 
linear combinations of the $n_{il}$ variables, which transform according 
to the irreducible representations $A_{1g}$, $A_{1u}$, $T_{1u}$, 
and $T_{2g}$ of the cubic group, 
\begin{equation}
Y_{A1g} = n_1 + n_2 + n_3 + n_4 + n_5 + n_6 + n_7 + n_8 \; ,
\label{A1g}
\end{equation}
\begin{equation}
Y_{A1u} = n_3 + n_4 + n_5 + n_7 - n_1 - n_2 - n_6 - n_8 \; ,
\label{A1u}
\end{equation}
\begin{subequations}
\begin{align}
Y_{T1u,1} = n_1 + n_2 + n_3 + n_4 - n_5 - n_6 - n_7 - n_8 \; , \\
Y_{T1u,2} = n_1 + n_2 + n_5 + n_6 - n_3 - n_4 - n_7 - n_8 \; , \\
Y_{T1u,3} = n_1 + n_4 + n_7 + n_8 - n_2 - n_3 - n_5 - n_6 \; , 
\label{T1u}
\end{align}
\end{subequations}
\begin{subequations}
\begin{align}
Y_{T2g,1} = n_1 + n_2 + n_7 + n_8 - n_3 - n_4 - n_5 - n_6 \; , \\
Y_{T2g,2} = n_1 + n_4 + n_5 + n_8 - n_2 - n_3 - n_6 - n_7 \; , \\
Y_{T2g,3} = n_1 + n_3 + n_6 + n_8 - n_2 - n_4 - n_5 - n_7 \; . 
\label{T2g}
\end{align}
\end{subequations}
Here we have omitted the cell index $i$. The variables $Y_{i\Gamma}$,
where $\Gamma = 1,2,\cdots ,8$ labels the symmetries in the 
above order, satisfy the relation
\begin{equation} 
Y_{i\Gamma}^2 = 1 \; .
\label{Y2}
\end{equation}
This implies, for example, that the $T_{1u}$ polar modes (3)
can be effectively represented by three independent Ising-type 
variables \cite{G1}.

We will assume that there exists a direct coupling between 
the symmetrized occupation probabilities or `pseudospins' $Y_{i\Gamma}$
due to long range dipolar interactions, which has the form
\begin{equation}
{\cal H}_{dir} = - \frac{1}{2}\sum_{i\ne j}\sum_{\Gamma}
I_{ij}^{\Gamma\Gamma'}Y_{i\Gamma}Y_{j\Gamma'}\; ,
\label{Hdir}
\end{equation}
In addition, we consider the interaction between the 
pseudospins and the phonon normal coordinates $Q_{\vec{q}p}$ \cite{G1,G2},
\begin{equation}
{\cal H}_{int} = - \frac{1}{\sqrt{N}}\sum_i\sum_{\vec{q}p\Gamma}
f_{\vec{q}p}^{\Gamma}\, Q_{\vec{q}p}\, Y_{i\Gamma}\, 
\exp({i\vec{q}\cdot\vec{R}_i})   \; ,
\label{Hint}
\end{equation}
where $\vec{q}$ is the wave vector and $p$ the branch 
index of lattice normal modes, $f_{\vec{q}p}^{\Gamma}$ the coupling 
constant, and $R_i$ the lattice vector. Introducing the phonon 
momenta $P_{\vec{q}p}$ and frequencies $\omega_{\vec{q}p}$, we 
can write down the phonon Hamiltonian \cite{B1}
\begin{equation} 
{\cal H}_{ph} = \frac{1}{2}\sum_{\vec{q}p}\left(\omega_{\vec{q}p}^2 
Q_{\vec{q}p}Q_{-\vec{q}p} + P_{\vec{q}p}P_{-\vec{q}p} \right) \; .
\label{Hph}
\end{equation}

It has long been established that in BaTiO$_3$ and other
perovskite ferroelectrics \cite{S3} a soft TO phonon mode exists. 
Here we assume that the soft TO mode has a wave vector $\vec{q}_0$ and 
symmetry $\Gamma =T_{1u,s}$, where $s$ is the corresponding branch index. 
The frequency of this soft mode vanishes at the 
stability limit $T_0$ in accordance with the Cochran relation
\begin{equation} 
\omega_{\vec{q}_0s}^2 = a(T - T_0) \; .
\label{omg0}
\end{equation}
In the following we will limit ourselves to the case where the soft mode 
condenses at the zone center, i.e., $\vec{q}_0 = 0$, as in BaTiO$_3$.

It is well known that the pseudospin and phonon degrees of freedom can 
be decoupled by introducing displaced phonon coordinates \cite{E1,G1,G2}
\begin{equation} 
{\tilde Q}_{\vec{q}p} = Q_{\vec{q}p} - \frac{1}{\sqrt{N}}\sum_{j\Gamma'} 
\frac{f_{-\vec{q}p}^{\Gamma'}}{\omega_{\vec{q}p}^2}\,Y_{\vec{q}\Gamma'} \; ,
\label{tQ}
\end{equation}
where
\begin{equation} 
Y_{\vec{q}\Gamma} = \sum_i Y_{i\Gamma} \exp(-i\vec{q}\cdot\vec{R}_i) \; .
\label{Yq}
\end{equation}
In this so-called adiabatic approach it is implied that the time scale for 
the Ti ion motion is much longer than the period of oscillation for phonon
modes, so that phonon coordinates $Q_{\vec{q}p}$ adapt adiabatically to 
any change of the pseudospin coordinates $Y_{i\Gamma}$. Thus we obtain the 
adiabatic Hamiltonian 
\begin{eqnarray} 
{\cal H}_{ad} = {\cal H}_{dir} + \frac{1}{2}\sum_{\vec{q}p}\left(
\omega_{\vec{q}p}^2 {\tilde Q}_{\vec{q}p}{\tilde Q}_{-\vec{q}p} + 
P_{\vec{q}p}P_{-\vec{q}p}\right) \nonumber\\
-\frac{1}{2}\sum_{ij} \sum_{\Gamma\Gamma'}K_{ij}^{\Gamma\Gamma'} 
\, Y_{i\Gamma}Y_{j\Gamma'} \;.
\label{Had}
\end{eqnarray}
In the last term, the Ti-Ti coupling constant is given by
\begin{equation} 
K_{ij}^{\Gamma\Gamma'} =  \frac{1}{N} \sum_{\vec{q}p}
\frac{f_{\vec{q}p}^{\Gamma}f_{-\vec{q}p}^{\Gamma'}}{\omega_{\vec{q}p}^2}\,
\exp[i\vec{q}\cdot(\vec{R}_i-\vec{R}_j)]  \; .
\label{Kij}
\end{equation}

On a mesoscopic scale, we are not interested in the microscopic mechanisms 
leading to the pseudospin-phonon coupling $f_{\vec{q}p}^{\Gamma}$ and 
merely adopt the established functional form \cite{G2}. 
In general, $f_{\vec{q}p}^{\Gamma}$ contains the contributions of 
both short- and long-range interactions between the Ti and all the other 
ions. Short-range interactions are expected to be primarily responsible 
for the local potential of the Ti ion. Indeed, first-principles calculations 
indicate that ferroelectricity in BaTiO$_3$ appears as the result of 
hybridization of the Ti-O bond \cite{C3,Z2}. Therefore, the contribution 
of short-range forces to $f_{\vec{q}p}^{\Gamma}$ and hence 
to $K_{ij}^{\Gamma\Gamma'}$ is expected to be dominant, although 
long-range electrostatic dipole-dipole interactions contained in 
$I_{ij}^{\Gamma\Gamma'}$ are also needed to establish ferroelectric 
order \cite{C3}. 

The $i \ne j$ part of the last term in Eq.~(\ref{Had}) has the same 
structure as the direct interaction (\ref{Hdir}). Therefore, 
$K_{ij}^{\Gamma\Gamma'}$ for $i \ne j$ represents an additional interaction 
between the Ti ions at two different sites, which together with the direct 
coupling $I_{ij}^{\Gamma\Gamma'}$ can lead to an order-disorder
transition into a $T_{1u}$ polarized state of the Ti subsystem. In general, 
this interaction involves two different symmetries $\Gamma$ and $\Gamma'$, 
and is real after being symmetrized with respect to the exchange 
$\Gamma \leftrightarrow \Gamma'$. Thus we can combine the two coupling 
constants $I_{ij}^{\Gamma\Gamma'}$ and $K_{ij}^{\Gamma\Gamma'}$
into a single coupling parameter $J_{ij}^{\Gamma\Gamma'} \equiv 
I_{ij}^{\Gamma\Gamma'} + K_{ij}^{\Gamma\Gamma'}$. 

The $i=j$ terms in Eq.\ (\ref{Kij}), however, represent a constant shift of 
the local energy and do not contribute to the ordering of Ti ions \cite{E1}. 
It can be shown by symmetry arguments that $K_{ii}^{\Gamma\Gamma'}$ is zero 
unless $\Gamma' = \Gamma$.  

In the following we will focus on the symmetric part of the interaction 
$K_{ij,\Gamma} \equiv K_{ij}^{\Gamma\Gamma}$. 
Its Fourier transform is given by 
\begin{equation} 
K_{\vec{q}\Gamma} = \sum_p \frac{\vert f_{\vec{q}p}^{\Gamma}\vert^2}
{\omega_{\vec{q}p}^2} - \frac{1}{N} 
\sum_{\vec{q'}p}\frac{\vert f_{\vec{q'}p}^{\Gamma}\vert^2}
{\omega_{\vec{q'}p}^2} \; .
\label{Kq}
\end{equation}
The second term is often neglected, however, as shown in Ref. \cite{E1}
its presence is crucial in order to ensure a zero average value of 
$K_{\vec{q}\Gamma}$. In the $\vec{q} \to 0$ limit, we
have $K_{0\Gamma} > 0$ if $\sum_p \vert f_{\vec{q}p}^{\Gamma}\vert^2/
\omega_{\vec{q}p}^2$ has a maximum at the zone center. For 
symmetry $\Gamma = T_{1u}$ this then favors a ferroelectric ordering
of the Ti subsystem provided that 
$J_{0\Gamma} \equiv I_{0\Gamma} + K_{0\Gamma} > 0$, 
implying that $\langle Y_{0T1u,s}\rangle \ne 0$,  ($s = 1, 2,$ or 3). 
If, however, the maximum occurs at the zone boundary, we can have 
$K_{0\Gamma} < 0$. For $J_{0\Gamma}< 0$, the ordering is antiferroelectric.
It should be stressed that in view of relation (\ref{omg0}) the interaction
$K_{\vec{q}\Gamma}$ is, in general, temperature dependent.

The off-diagonal coupling $J_{\vec{q}}^{\Gamma\Gamma'}$ 
with $\Gamma' \ne \Gamma$ leads to anisotropic 
interactions, which are assumed to be weaker than the isotropic
part and can thus be treated by perturbation theory. Below, we will 
discuss the possibility that anisotropic interactions give rise to 
time-dependent random fields acting on the pseudospin variables.

In a ferroelectric system like BaTiO$_3$, the main contribution to the
coupling $K_{\vec{q}\Gamma}$ will come from the soft TO mode with phonon
coordinate ${\tilde Q}_{0s}$ and frequency $\omega_{0s}$ as given by 
Eq.~(\ref{omg0}). The Ti subsystem will undergo a phase transition into 
an ordered state with nonzero value of the pseudospin thermal average 
$\langle Y_{0s}\rangle \ne 0$. The transition temperature is determined 
by the relation
\begin{equation} 
kT_c = I_{s0} + \frac{f_{0s}^2}{a(T_c-T_0)} - L_{s0} \; ,
\label{kTc}
\end{equation}
where we have substituted expression (\ref{omg0}) for $\omega_{0s}^2$ 
and written 
$L_{s0} \equiv (1/N)\sum_{\vec{q'}} 
\vert f_{\vec{q'}s}\vert^2/\omega_{\vec{q'}s}^2$.
To evaluate $L_{s0}$, we should know the details of the phonon spectrum as well 
as the $\vec{q}$-dependence of the coupling. By adding a $q^2$-term to 
$\omega_{0s}$ in Eq.~(\ref{omg0}), it can be shown that the leading 
contribution to $L_{s0}$ is a constant independent of temperature. Thus
we find
\begin{equation} 
T_c = \frac{1}{2}\left[T_0+M_0/k + \sqrt{(T_0-M_0/k)^2 
+ 4f_0^2/ka}\,\right] \; ,
\label{Tc}
\end{equation}
where $M_0 \equiv I_0 - L_0$ and we have dropped the indices $s$. 
Expression (\ref{Tc}) differs from the result of GY by the presence 
of the $M_0$-term, which is, in general, different from zero. We will
assume that $I_0 > L_0$ and $0 < M_0 < T_0$, implying that $T_c > T_0$, i.e.,                                                                                                                           
the ordering takes place above the stability limit of the unperturbed soft 
mode. It can easily be seen that in the above case, the effect of $M_0$ 
is to shift $T_c$ towards higher temperatures.


\section{Soft mode dynamics}

We now consider the response of the soft TO phonon mode $Q_{\vec{q}s}$ to a 
time-dependent electric field $E_{-\vec{q}s}$ associated with an optic 
wave in a light scattering experiment. Close to the transition, the time
scales for the soft mode and for the relaxational motion of the Ti ions
become comparable, and the dynamics based on the adiabatic Hamiltonian 
(\ref{Had}) is not applicable. The soft mode dynamics is governed by the 
corresponding part of the original Hamiltonian, namely,
\begin{eqnarray} 
&{\cal H}_{sm}  =  - \frac{1}{2} \sum_{\vec{q'}} I_{\vec{q'}s} Y_{\vec{q'}s}
Y_{-\vec{q'}s} \nonumber\\
&+ \frac{1}{2}\sum_{\vec{q'}} \left( \omega_{\vec{q'}s}^2
Q_{\vec{q'}s}Q_{-\vec{q'}s} + P_{\vec{q'}s}P_{-\vec{q'}s}\right)\nonumber\\ 
~~\nonumber\\
&- \frac{1}{\sqrt{N}} \sum_{\vec{q'}} f_{\vec{q'}s} Q_{\vec{q'}s} Y_{-\vec{q'}s} 
- \mu E_{-\vec{q}s} Q_{\vec{q}s}\exp(i\omega t) \; . 
\end{eqnarray}
\label{Hsm}
Here $\mu = e^*/\sqrt{m^*}$ is a coupling parameter 
involving the effective charge $e^*$ and reduced mass $m^*$. For
simplicity, we do not include the direct coupling between the light
vector $E_{-\vec{q}s}$ and the dipole moment associated with the Ti-O
bond. We could, in principle, assume a nonzero value of the off-center Ti
dipole moment, however, as shown by GY its contribution to the Curie
constant is negligible in other perovskite ferroelectrics and can thus 
be ignored.  

The time evolution of the soft-mode operators $Q_{\vec{q}s}$ and 
$P_{\vec{q}s}$ is governed by the Heisenberg equations of motion 
$dQ_{\vec{q}s}/dt = -(i/\hbar)[Q_{\vec{q}s}, {\cal H}_{sm}]$. 
In contrast, the time dependent thermodynamic fluctuations 
of the variables $Y_{is}(t)$ will be assumed to exhibit a pure relaxational 
motion with a single characteristic relaxation time $\tau$,
i.e., we will ignore the possibility of coherent dipole moment flips. 
The corresponding equation of motion can be obtained from the classical 
Langevin model \cite{FH}, which is based on the continuous or `soft' spin 
variables $-\infty < Y_{is}(t) < +\infty$ with effective Hamiltonian
\begin{equation} 
\beta {\cal H}_{eff} = \beta{\cal H}_{sm} + \sum_i\left(\frac{1}{2}\,
r Y_{is}^2 + \frac{1}{4}\, u Y_{is}^4 \right) \; .
 \label{LK}
\end{equation}
For $u = -r \to \infty$ one recovers the discrete limit $Y_{is}^2 = 1$ 
(cf. Eq.~(\ref{Y2})). The Langevin equation of motion is
\begin{equation} 
\tau \frac{\partial Y_{is}}{\partial t} = - \frac{\partial\beta({\cal H}_{eff})}
{\partial Y_{is}} + \xi_{\vec{q}s}(t) \; ,
\label{LE}
\end{equation}
where the Langevin noise $\xi_{is}(t)$ is a Gaussian random variable with zero 
mean and variance
\begin{equation} 
\langle \xi_{is}(t)\xi_{js}(t')\rangle = 2\tau \delta_{ij}\delta(t-t') \; .
\label{xi}
\end{equation}
Introducing the Fourier components $Y_{\vec{q}s}(\omega )$ etc., 
we obtain the linearized equations of motion 
\begin{subequations} 
\begin{align}
i\omega Q_{\vec{q}s} &= P_{\vec{q}s} \; ;\\
i\omega P_{\vec{q}s} &= - \omega_{\vec{q}s}^2 Q_{\vec{q}s}
+\frac{1}{\sqrt{N}} f_{-\vec{q}s} Y_{\vec{q}s} + \mu E_{\vec{q}} \; ; \\
i\omega\tau Y_{\vec{q}s} &= - (r - \beta I_{\vec{q}s} - \Sigma_{\vec{q}s}) 
Y_{\vec{q}s} + \beta \sqrt{N} f_{\vec{q}s}Q_{\vec{q}s} + \xi_{\vec{q}s}(\omega)\;. 
\label{eqm}
\end{align}
\end{subequations}
Here $\Sigma_{\vec{q}s}$ is the self energy which can, in principle, be 
calculated by a diagrammatic expansion involving the parameter $u$ and
the pseudospin-phonon coupling $f_{\vec{q}s}$. In the following 
we will ignore the frequency dependence of $\Sigma_{\vec{q}s}$ in the 
soft-mode regime.

We can now introduce the static pseudospin response 
$\chi_{\vec{q}s} = \langle\delta Y_{\vec{q}s}/\delta
\xi_{\vec{q}s}\rangle$, i.e., 
\begin{equation} 
\chi_{\vec{q}s} = \frac{\beta}{r - \beta I_{\vec{q}s}-\Sigma_{\vec{q}s}} \; ,
\label{chi}
\end{equation}
and redefine the relaxation time by writing 
$\tau_{\vec{q}s} \equiv \tau \chi_{\vec{q}s}/\beta$. The solution of the above 
equations can then be expressed in the form 
\begin{equation} 
Q_{\vec{q}s}(\omega) = \frac{1}{\omega_{\vec{q}s}^2 - \omega^2 
-\vert f_{\vec{q}s}\vert^2\chi_{\vec{q}s}/(1 + i\omega\tau_{\vec{q}s})}
\, \mu E_{\vec{q}}\; ,
\label{QE}
\end{equation}
and 
\begin{equation} 
Y_{\vec{q}s}(\omega) = \frac{\beta\sqrt{N}f_{\vec{q}s}\chi_{\vec{q}s}}
{1 + i\omega \tau_{\vec{qs}}}Q_{qs}(\omega)  \; ,
\label{YQ}
\end{equation}
where the random force term has been averaged out.  
Eq.~(\ref{QE}) is similar, but not identical to the expression given by GY, 
which contains an extra term $i\omega\Gamma_{qs}$ in the denominator, 
describing the damping of the TO soft mode due to lattice anharmonicity. 
Here we absorb this damping into the $i\omega\tau_{\vec{q}s}$ term by  
a proper redefinition the relaxation time $\tau_{\vec{q}s}$. 

We now consider the $\vec{q} = 0$ case corresponding to a TO soft
mode at the zone center. Dropping the subscripts $s$ and introducing
the far-infrared dielectric response $\chi_Q(\omega ) = 
Q_0(\omega)/(\mu E_0)$ we get 
\begin{equation} 
\chi_Q(\omega ) = \frac{1+i\omega\tau_0}{(\omega_0^2-\omega^2)
(1+i\omega\tau_0) - f_0^2\chi_0} \; .
\label{cQ}
\end{equation}
The contribution to the corresponding dielectric function is given by
\begin{equation} 
\epsilon_Q (\omega ) \simeq \epsilon_{\infty} 
+ \frac{\mu^2}{\epsilon_0 v_0}\, \chi_Q(\omega ) \; ,
\label{eps}
\end{equation}
where $v_0$ is the unit cell volume.

In the absence of pseudospin-phonon coupling, $f_0 \to 0$, the static 
response $\chi_Q(0)$ diverges at $T_c = T_0$ in view of 
Eq.~(\ref{omg0}). For $f_0 \ne 0$, the critical temperature is obtained from 
the equation $\omega_0^2 - f_0^2\chi_0 = 0$, or after applying Eq.~(\ref{chi}) 
from the relation 
\begin{equation} 
a(T - T_0) = \frac{\beta f_0^2}{(r - \beta I_0 - \Sigma_0)} \; .
\label{af}
\end{equation}
Returning to the discrete limit $Y_{is}^2 = 1$ we notice that the last
result will be equivalent to Eq.~(\ref{kTc}) provided that 
$r - \Sigma_0 \simeq 1 + \beta L_0$. Thus, in the above limit the 
critical temperature $T_c$ derived from the soft-mode dynamics will be 
precisely equal to the static value (\ref{Tc}). This means that the 
hybrid soft mode becomes unstable at the static ordering temperature 
$T_c$ of the Ti subsystem. It should be added that the technical reason 
for using the soft spin variables and Langevin dynamics is the fact that 
only in this formalism we were able to properly take into account the
presence of the second term in Eq.~(\ref{Kq}).

To illustrate the frequency dependence of $\chi_Q(\omega )$ we 
introduce dimensionless parameters $f_0 \to f_0\tau_0/\sqrt{kT_0}$ and 
$a \to aT_0\tau_0^2$, and choose their representative values as  
$f_0 = 0.4$ and $a = 0.145$, respectively. Also, we fix the value of the 
parameter $M_0$ at $M_0/kT_0 = 0.5$ and rescale the frequencies as  
$\omega \to \omega\tau_0$. In all numerical calculations we furthermore 
set $\tau_0$~=~const., i.e., we ignore the non-critical temperature 
dependence of the relaxation time. We can then evaluate the spectral function
\begin{equation} 
I_Q(\omega) = \frac{\chi_Q''(\omega)}{\omega}  \; ,
\label{Iom}
\end{equation}
which is related to the Raman scattering intensity. Here the imaginary 
part of the response (\ref{cQ}) is defined by 
$\chi_Q(\omega ) = \chi_Q'(\omega ) - i\chi_Q''(\omega )$. In Fig.~1,  
we plot $I_Q(\omega)$ at five different temperatures $T/T_c$. 
Far from $T_c$, the spectrum is characterized by 
a single peak, the position of which moves towards zero frequency as 
$T \to T_c$ in accordance with the expected soft-mode behavior. 

The hybrid soft-mode frequency approximately corresponds to the
position of the peak in $I_Q(\omega)$. 
To obtain an unambiguous definition it is necessary to consider 
the complex poles of the response function (\ref{chi}). After introducing a 
new variable $z = i\omega\tau_0$, the denominator takes the
standard form of a cubic polynomial with real coefficients
\begin{eqnarray} 
z^3 + z^2 +\omega_0^2\tau^2\,z+ (\omega_0^2 - f_0^2\chi_0)\tau_0^2 \nonumber\\
= (z - z_1)(z - z_2)(z - z_3) = 0 \; .
\label{z}
\end{eqnarray}
The complex zeros $z_n$, $n = 1,2,3$, are given by the Cardano 
formula. The corresponding analytic expressions for 
$z_n$ are cumbersome, but can easily be evaluated numerically. 
They then lead to three complex frequencies 
$\omega_n = x_n + iy_n$ having the following properties: 
(i) One of the frequencies is always imaginary, say, $\omega_1 = iy_1$; 
(ii) The remaining two solutions are, in general, complex
and such that always $x_3 = x_2$; (iii) If $x_2 = 0$, then $y_3 \ne y_2$, 
however, for $x_2 \ne 0$ we have $y_3 = y_2$; 
(iv) In general, $\sum_n y_n = 1/\tau_0$, and (v) $\omega_1\omega_2 +
\omega_2\omega_3 + \omega_3\omega_1 = -\omega_0^2$.

The response (\ref{cQ}) thus becomes
\begin{equation} 
\chi_Q(\omega ) = - \frac{\omega - i\nu_0}
{(\omega - iy_1)(\omega^2 - \Omega_0^2 -i\omega\gamma)} \; ,
\label{c1}
\end{equation}
where $\nu_0 \equiv 1/\tau_0$ and the resonance frequency $\Omega_0$ 
is given by
\begin{equation} 
\Omega_0^2 = - \omega_2 \omega_3 = x_2^2 + y_2 y_3 \;  ,
\label{om2}
\end{equation}
while the damping parameter is
\begin{equation} 
\gamma = y_2 + y_3 .
\label{gam2}
\end{equation}
The term $i(x_2y_3 + x_3y_2)$ formally appearing in the product 
$\omega_2 \omega_3$ is zero in view of the properties (ii) and 
(iii) above. It further follows from (iv) that $y_1 = \nu_0 - \gamma$, 
whereas from (v) we obtain a relation between $\Omega_0$ and $\gamma$, 
namely, 
\begin{equation} 
\Omega_0^2 + \gamma(\nu_0 - \gamma) =\omega_0^2 \; .
\label{Og}
\end{equation}

We can now rewrite the spectral function (\ref{Iom}) in the form
\begin{equation} 
I_Q(\omega ) =\frac{A}{\omega^2 + y_1^2}
+ \frac{B - A\omega^2}{(\omega^2 - \Omega_0^2)^2 + \omega^2\gamma^2} \; ,
\label{Iomr}
\end{equation}
where $A = \gamma/(\Omega_0^2+y_1^2-y_1\gamma )$ and 
$B=(2\Omega_0^2+y_1^2-\gamma^2)A$. The first term represents a central peak 
associated with the relaxational motion of the Ti ions, whereas the second 
term describes a resonance occurring at $\Omega_0$. It should be noted,
however, that because of the $-A\omega^2$ term in the numerator the last 
expression deviates from the standard spectral function for the damped 
harmonic oscillator. Furthermore, the $-A\omega^2$ term partially compensates
the contribution of the relaxational term in Eq.~(\ref{Iomr}), so that a 
separate central peak cannot be discerned at temperatures of interest, 
as it is evident from Fig.~1. 

In Fig.~2, we plot $\Omega_0^2$ as a function of temperature with
the same parameter values as in Fig.~1. For $f_0$ smaller than
a limiting value $f_0^m$, where $f_0^m= 0.6721$ for the present choice
of representative parameter values, the temperature dependence of 
$\Omega_0^2$ near $T_c$ is quasi linear, 
\begin{equation} 
\Omega_0^2 \simeq a'(T - T_c) \; ,
\label{om2T}
\end{equation}
where $a' = a'(f_0)$ with $a' \ge a$ and $T_c \ge T_0$.
Thus we are dealing with a hybrid soft mode, which condenses at the 
instability temperature $T_c$ given by the static critical temperature
(\ref{kTc}).  

The temperature dependence of the damping constant $\gamma$ 
is shown in Fig.~3. For $f_0 < f_0^m$ we have $0 < \gamma < \nu_0$,
whereas for $f_0 > f_0^m$ we can see that $\gamma \to \nu_0$ as $T \to T_c$. 
It can be shown that in the hybrid soft mode regime $f_0 < f_0^m$ the 
reduced damping constant $\gamma_0 = \gamma/\Omega_0$ diverges on 
approaching $T_c$ from above as $\gamma_0 \sim (T - T_c)^{-1/2}$. 
The temperature dependence of $\gamma_0$ is shown in the inset of Fig.~3. 

The static susceptibility as obtained from Eq.~(\ref{c1}) is given by
\begin{equation} 
\chi_Q(0) = \frac{1}{\tau_0\Omega_0^2(\nu_0 - \gamma)} \; ,
\label{chis}
\end{equation}
and is found to diverge at $T_c$ as $\chi_Q(0) \sim (T - T_c)^{-1}$ for all 
values of $f_0$. As already stated, for $f_0 > f_0^m$ the damping parameter  
$\gamma$ approaches the value $\gamma \to \nu_0$ as $T \to T_c$,
and thus $\chi_Q(0)$ diverges even for $\Omega_0(T_c) \ne 0$. The
temperature dependence of the inverse static response $\chi_Q^{-1}$ 
is displayed in Fig.~2. For $T < T_c$, we formally have $\Omega_0^2 < 0$ and
the system becomes unstable. In principle, it should be stabilized by
including anharmonic terms into the phonon Hamiltonian (\ref{Hph}) 
as well as a coupling to elastic strains.

For $f_0 > f_0^m$, ${\Omega}_0$ is always nonzero and strictly speaking 
the hybrid soft-mode concept is not applicable in that case. The corresponding 
limiting value of the instability temperature corresponding to the 
representative parameter values used in numerical calculations here 
is $T_c^m/T_0 = 2.533$. This is a fairly large value, suggesting that 
any real system is likely to be in the soft-mode regime $0 < f_0 < f_0^m$.


\begin{figure}
\begin{center}
\epsfig{file=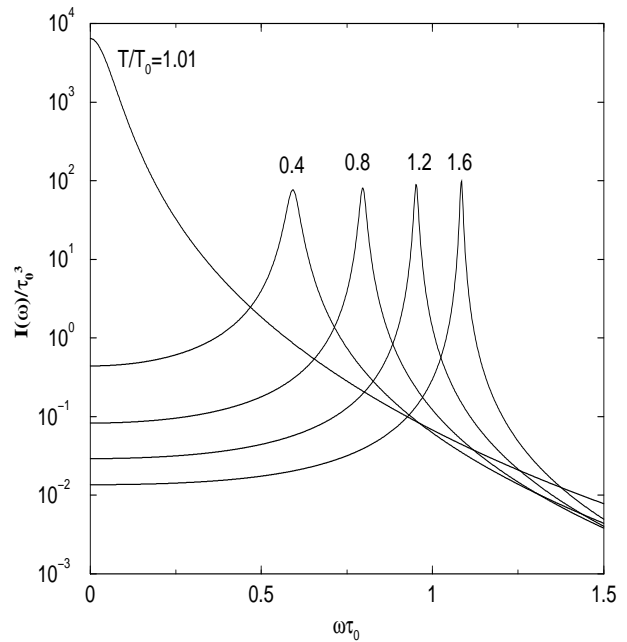,height=3.5in,width=3.2in}
\end{center}
\caption{Spectral density of far-infrared dielectric response,
plotted on a logarithmic vertical scale,  
calculated for pseudospin-phonon coupling strength $f_0 = 0.4$ and
for several values of temperature, as indicated.}
\label{fig1}
\end{figure}

\begin{figure}
\begin{center}
\epsfig{file=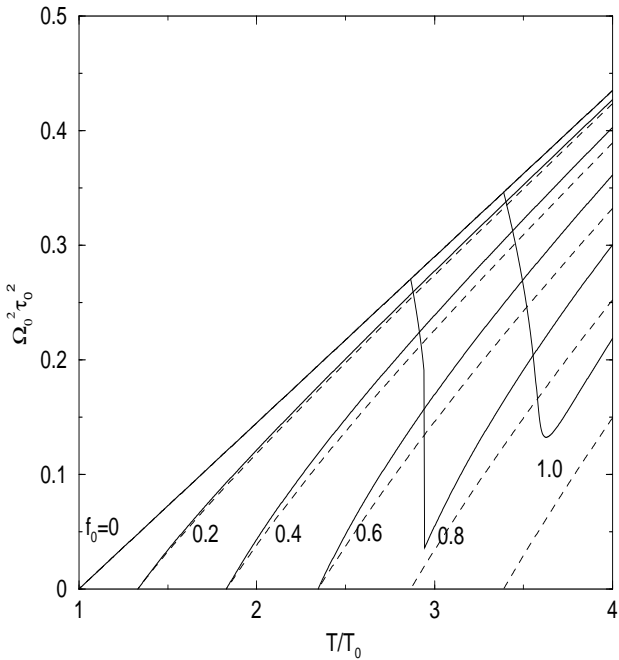,height=3.5in,width=3.2in}
\end{center}
\caption{Resonance frequency $\Omega_0$ as a function of temperature
for several values of coupling strength $f_0$ (solid lines).
For $f_0 < f_0^m = 0.6721$ a soft-mode behavior is observed.
Also shown is the temperature dependence of the inverse static 
response (dashed lines).}
\label{fig2}
\end{figure}

\begin{figure}
\begin{center}
\epsfig{file=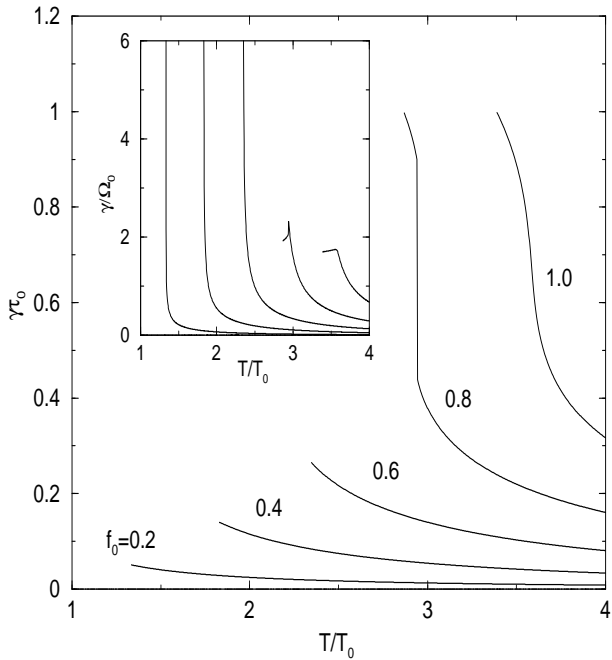,height=3.5in,width=3.2in}
\end{center}
\caption{Damping parameter $\gamma$ as a function of temperature
for several values of coupling strength $f_0$. Inset: Reduced 
damping parameter $\gamma_0 = \gamma/\Omega_0$ for the same values
of $f_0$.}
\label{fig3}
\end{figure}

\section{Discussion}

To determine the type of the phase trasition occurring at
$T = T_c$, we introduce the order parameter as the thermal average
of the Ti pseudospin variable $P = \langle Y_0 \rangle$, which is related 
to $\langle Q_0 \rangle$ through Eq.~(\ref{YQ}) at $\omega = 0$.
We can then write down a Landau-type free energy
\begin{equation} 
{\cal F}(P) = \frac{1}{2}\,\chi_Y(0)^{-1}P^2 + 
\frac{1}{4}\,b\, P^4 + \frac{1}{6}\, c P^6 + \cdots  \; ,
\label{FL}
\end{equation}
where the static pseudospin response $\chi_Y(0)$ is related to the hybrid 
soft-mode response (\ref{cQ}) and the rigid static pseudospin 
response (\ref{chi}), namely, 
\begin{equation} 
\chi_Y(0) = \chi_Q(0) \frac{\omega_0^2}{\omega_0^2 - f_0^2\chi_0}
\, \left(\frac{\omega_0^2}{f_0^2\chi_0}\right) \; . 
\label{cY}
\end{equation}
As noted in deriving Eq.~(\ref{af}), the first denominator vanishes as 
$T \to T_c$, while in the same limit the last factor approaches the value 
$\omega_0^2/f_0^2\chi_0 \to 1$. Eq.~(\ref{cY}) is reminiscent of the
Lyddane-Sachs-Teller relation for the static dielectric response. It
follows that the coefficient of the quadratic term in ${\cal F}(P)$
vanishes at $T_c$ as $\sim (T - T_c)$. The coefficients $b$ and $c$ can
easily be determined within a molecular field approximation, however, it
turns out that this approximation would predict a second order phase 
transition, in disagreement with observations in BaTiO$_3$. To derive  
the correct value of the coefficient $b$ in the free energy we would have
to include, for example, a coupling between the soft mode and elastic 
strain \cite{H1}, or equivalently, between the strain and the Ti pseudospin 
variable. The appearance of a nonzero value of the order parameter will 
thus be accompanied by a macroscopic deformation of the lattice with 
tetragonal symmetry. In practice, $b$ and $c$ are often considered as
phenomenological parameters. Hatta and Ikushima \cite{H2} applied an 
expression of the above form to analyze the measured heat capacity 
of BaTiO$_3$ at constant electric field. They used the value 
$T_c = 383$ K (in our notation) and found that $b < 0$ and $c > 0$, 
implying a first order phase transition which takes place at 
$T_C \simeq 398$ K (or at $\simeq 408$ K, 
depending on the sample preparation method). They also determined 
the jump in the heat capacity at $T_C$ of the order 
$\Delta C \simeq 0.19$ k, from which it was concluded that the Ti ion 
lies in a single minimum potential, in apparent disagreement with 
the off-center Ti picture. According to the present model the number of 
equilibrium positions of Ti is greater than one, but their actual number 
is only relevant for the definition of the Ti pseudospin variables. 
If we restrict the discussion to the adiabatic Hamiltonian (\ref{Had}), 
the phase transition appears to be of order-disorder type. However, the 
above approach involving hybrid soft-mode dynamics---which leads to 
the same critical temperature as the static approach---clearly has the
characteristics of a displacive transition, but with a simultaneous ordering 
of the Ti subsystem as an additional order-disorder feature.

Evidence of a TO soft mode with $T_{1u}$ symmetry in BaTiO$_3$ has been 
found by Vogt et al. \cite{V1} using the hyper-Raman scattering technique.
The dielectric function has been fitted by a classical single oscillator 
dispersion formula, which is equivalent to the second term of Eq.~(\ref{Iomr}) 
without the $\omega^2$-term in the numerator. One of the important results 
of  Ref.~\cite{V1} is that the frequency of the soft mode tends to zero 
close to the transition temperature and no saturation 
of the frequency at a finite value reported earlier occurs \cite{L1}, 
in agreement with the predictions of the present model.
Close to $T_c$, the temperature dependence of the soft-mode frequency 
can be described by the relation (\ref{om2T}) with 
$a' \simeq 1.4 \times 10^{22}$ s$^{-2}$K$^{-1}$. The relative damping 
constant $\gamma/\Omega_0$ was found to exceed the value of 2 at all 
temperatures, whereas in the present model this is true only close 
to $T_c$.

It has been shown by NMR methods \cite{Z1} that in the high temperature 
phase of BaTiO$_3$ the unit cells are tetragonally distorted, although the 
overall macroscopic symmetry is cubic. This agrees with Wada et al. 
\cite{W1} who determined the symmetry by Raman scattering and found that 
the {\it microscopic} symmetry was P4mm both above and below $T_c$. Since 
the orientation of the tetragonal axis varies across the crystal, the
{\it macroscopic} symmetry is Pm3m above the transition and P4mm below $T_C$.
This can be accounted for by the present model if we return to the Ti-Ti
coupling terms (\ref{Hdir}) and (\ref{Kij}), and allow for off-diagonal 
interactions with $\Gamma' \ne \Gamma$. This then gives rise to extra terms 
the equations of motion (\ref{eqm}) for $Y_{\vec{q}s}$, e.g.,
\begin{eqnarray} 
i\omega\tau Y_{\vec{q}\Gamma} =& - (r - \beta I_{\vec{q}\Gamma} - 
\Sigma_{\vec{q}\Gamma}) Y_{\vec{q}\Gamma} + 
\beta \sqrt{N} f_{\vec{q}\Gamma}Q_{\vec{q}\Gamma} \nonumber\\ 
&+ \beta \sum_{\Gamma'\ne \Gamma}J_{\vec{q}}^{\Gamma\Gamma'}Y_{\vec{q}\Gamma'} 
+ \xi_{\vec{q}\Gamma}(\omega) \; .
\label{eqm2}
\end{eqnarray}
On the time scale of the TO soft mode and of the Ti relaxational mode 
$Y_{\vec{q}\Gamma} = Y_{\vec{q}s}$ the variables $Y_{\vec{q}\Gamma'}$
are much slower and appear to be "frozen" in a given configuration. Thus 
the sum over $\Gamma'$ will play the role of a random variable
$h_{\vec{q}\Gamma} = \sum_{\Gamma'\ne \Gamma}
J_{\vec{q}}^{\Gamma\Gamma'}Y_{\vec{q}\Gamma'}(t)$ analogous to the random 
electric field in dipolar glasses. Instead of spatial randomness, however, 
we are dealing here with a temporal disorder, which appears to be 
"quenched" on the time scale of $Y_{\vec{q}\Gamma}(t)$. This field will give 
rise to a slowly varying deformation of $T_{1u,s}$ symmetry, which
is experimentally observable in both NMR and Raman experiments. Since 
the orientation of the deformation axis $s$ varies both in space and
time, the average symmetry of the system remains cubic.


\section{Conclusions}

Recent experiments provided new evidence that the Ti ion in the
cubic phase of BaTiO$_3$ \cite{Z1} and other oxide ferroelectrics \cite{S4}
occupies one of the eight off-center positions in the unit cell along 
the $[111]$ directions. We have presented a simple mesoscopic model 
of BaTiO$_3$, assuming that the symmetrized occupational probabilities for 
the Ti sites or pseudospins are linearly coupled with the normal coordinates
of lattice vibrations. On the time scale of Ti intersite jumps, lattice modes 
act as fast variables, which can adapt instantaneously to any change of 
the pseudospin configuration. In this adiabatic approximation, the 
pseudospin-phonon coupling gives rise to a static phonon mediated effective 
Ti-Ti interaction. The Ti subsystem thus undergoes an order-disorder transition 
into a polarized state of $T_{1u}$ symmetry, and the critical temperature 
$T_c$ is determined by a sum of the contributions from the direct Ti-Ti 
coupling and the pseudospin-lattice part. The leading contribution is due 
to the TO soft mode, the frequency of which tends to zero at 
the instability temperature $T_0$. In general, $T_c \ge T_0$, where 
the shift of $T_c$ depends both on the direct interaction as well as 
the pseudospin-phonon coupling. 

Close to the stability limit for he TO soft mode, the time scales for 
the Ti and lattice motion become comparable and a dynamic treatment becomes 
necessary. By combining the classical Langevin equations of motion for
the relaxation of the Ti pseudospins and the quantum Heisenberg equations
for the TO soft mode coordinate, we have obtained the response of the system 
to an oscillating electric field, typically in an optical scattering
experiment. The spectral function of this response consists of a central 
peak and a resonance, belonging to a new hybrid soft mode with $T_{1u}$ 
symmetry, which describes the in-phase motion of the TO soft optic
mode and the relaxational decay of the symmetrized Ti pseudospin variable.
The frequency of the hybrid soft mode tends to zero at the static critical 
temperature $T_c$. The central peak component is determined by the 
relaxational motion of the Ti subsystem and can be regarded as a typical
order-disorder feature of the total response, in contrast to the resonance 
which shows a mixed displacive and order-disorder character.

The TO soft mode in BaTiO$_3$ has been observed experimentally by hyper-Raman
scattering \cite{V1}, and has been described by a damped harmonic oscillator,
similar to our resonance term in the dynamic response. 
The soft mode frequency was found to extrapolate to zero close to the
thermodynamic transition temperature $T_C$.

The frequency of the hybrid soft mode depends on the Ti-phonon coupling
strength and exists only in a finite interval $0 < f_0 < f_0^m$, where 
the value of $f_0^m$ depends on the model parameters. Outside 
this interval, the soft mode concept is not applicable, however, the
static response always diverges at $T_c$. 

To calculate the free energy of the system we would have to include a 
coupling between the soft mode and lattice strains into the model.
This can be done in a phenomenological approach \cite{H1}, which describes 
the first order phase transition from the high temperature phase with 
macroscopic cubic symmetry to a tetragonal low temperature phase. 
The corresponding Curie temperature $T_C $ is generally higher than 
the instability temperature $T_c$ associated with the condensation 
of the hybrid soft mode.

\section{Acknowledgment}

This work was supported by the Ministry of Education, Science, and Sport 
of The Republic of Slovenia. The authors are grateful to Ekhard K.~H. Salje 
for useful discussions.





\begin{thebibliography}{}
\bibitem{S1} G. Shirane, B. C. Frazer, V. J. Minkiewicz, and J. A. Leake,
Phys. Rev. Lett. {\bf 19}, 234 (1967).
\bibitem{V1} H. Vogt, J. A. Sanjurjo, and G. Rossbroich,
Phys. Rev. B {\bf 26}, 5904 (1982).
\bibitem{C1} R. Comes, M. Lambert, and A. Guinier, 
Solid State Commun. {\bf 6}, 715 (1976).
\bibitem{C2} A. S. Chaves, F. C. S. Barreto, R. A. Nogueira, and B. Zeks,
Phys. Rev. B {\bf 13}, 207 (1982).
\bibitem{Z1} B. Zalar, V. V. Laguta, and R. Blinc,
Phys. Rev. Lett. {\bf 90}, 037601 (2003).
\bibitem{C3} R. E. Cohen, Nature (London) {\bf 358}, 136 (1992).
\bibitem{A1} S. Aubry, J. Chem. Phys. {\bf 62}, 3217 (1975);
{\it ibid.} {\bf 64}, 3392 (1976).
\bibitem{S2} M. Stachiotti, A. Dobry, R. Migoni, and A. Bussmann-Holder,
Phys. Rev. B {\bf 47}, 2473 (1993).
\bibitem{G1} Y. Girshberg and Y. Yacoby,
Solid State Commun. {\bf 103}, 425 (1997). 
\bibitem{G2} Y. Girshberg and Y. Yacoby,
J. Phys.: Condens. Matter {\bf 11}, 9807 (1999).
\bibitem{B1} R. Blinc and B. \v Zek\v s, {\it Soft Modes in Ferroelectrics
and Antiferroelectrics} (North Holland, Amsterdam, 1974).
\bibitem{S3} For a review, see: J. F. Scott, Rev. Mod. Phys. {\bf 46},
83 (1974).
\bibitem{E1} R. J. Elliott, G. A. Gehring, A. P. Malozemoff, S. R. P. Smith, 
W. S. Staude, and R. N. Tyte, J. Phys. C {\bf 4}, L179 (1971).
\bibitem{Z2} W. Zhong, D. Vanderbilt, and K. M. Rabe, 
Phys. Rev. B {\bf 52}, 6301 (1995).
\bibitem{FH} K. H. Fischer and J. A. Hertz, {\it Spin Glasses}
(Cambridge University Press, Cambridge, UK, 1991), p. 131.
\bibitem{H1} S. A. Hayward and E. K. H. Salje, 
J. Phys.: Condens. Matter {\bf 14}, L599 (2002).
\bibitem{H2} I. Hatta and A. Ikushima, J. Phys. Soc. Japan {\bf 41},
558 (1976). 
\bibitem{L1} Y. Luspin, J. L. Servoin, and F. Gervais,
J. Phys. C {\bf 13}, 3761 (1980).
\bibitem{W1} S. Wada, T. Suzuki, M. Osada, M. Kakihana, and T. Noma,
Jpn. J. Appl. Phys. {\bf 37}, 5385 (1998).
\bibitem{S4} N. Sicron, B. Ravel, Y. Yacoby, E. A. Stern, F. Dogan,
and J. J. Rehr, Phys. Rev. B {\bf 50}, 13168 (1994).
\end{thebibliography}
\end{document}